\documentclass[journal=jctcce,manuscript=article,layout=twocolumn]{achemso}
\setkeys{acs}{doi=true}

\usepackage{graphicx}
\usepackage{dcolumn}
\usepackage{bm}
\usepackage{ulem}
\usepackage{xcolor}
\usepackage{color,soul}
\usepackage{amsmath}
\usepackage{hyperref}
\hypersetup{colorlinks,allcolors=blue}
\DeclareUnicodeCharacter{0308}{\"{u}}

\title{
Water Phase Diagram from a General-Purpose Atomic Cluster Expansion Potential
}

\author{Eslam Ibrahim}
\email{eslam.saadibrahim@rub.de}
\affiliation{ICAMS, Ruhr Universit\"at Bochum, 44780 Bochum, Germany}
\author{Yury Lysogorskiy}
\author{Ralf Drautz}
\email{ralf.drautz@rub.de}
\affiliation{ICAMS, Ruhr Universit\"at Bochum, 44780 Bochum, Germany}
\author{Pablo M. Piaggi}
\email{pm.piaggi@nanogune.eu}
\affiliation{CIC nanoGUNE, Tolosa Hiribidea 76, Donostia 20018, San Sebastian, Spain}
\alsoaffiliation{Ikerbasque, Basque Foundation for Science, Bilbao 48013, Spain}

\begin{document}

\begin{abstract}
Water’s phase diagram remains one of the most intricate and challenging benchmarks in molecular modeling. 
In this study, we compute the phase diagram of water using an Atomic Cluster Expansion (ACE) potential trained on density-functional theory (DFT) calculations based on the revPBE-D3 exchange and correlation functional.
We compute solid–liquid chemical potential differences and melting points using biased coexistence simulations with the On-the-Fly Probability Enhanced Sampling (OPES) method.
Starting from these points, we trace coexistence lines using Gibbs–Duhem integration.
This combination of methods allows us to consistently map pressure–temperature phase boundaries and reconstruct the full phase diagram between approx.~100--500\,K and 0--4\,GPa.
The stability regions of the main ice polymorphs (Ih, II, V, VI, and VII) are reproduced in close agreement with experiments.
As in earlier studies based on DFT, ice III is metastable and there are systematic shifts of coexistence lines with respect to experimental results. 
Our results demonstrate the capability of our general-purpose ACE potential to capture the complex phase behavior of water across wide thermodynamic conditions.
\end{abstract}

\maketitle

\section{Introduction}

Water exhibits a remarkable range of physical and structural properties that continue to challenge theoretical and experimental understanding.
Its complex chemical bonding gives rise to anomalies, such as the density maximum at 4\,°C and the high heat capacity, and to a vast polymorphism in its crystalline and amorphous phases~\cite{head2002water,stillinger1980water,debenedetti2003supercooled}.
In the temperature range up to about 400\,K and pressures below 50\,GPa, water displays a rich phase behavior, showing at least nine stable crystalline ice phases (Ih, II, III, V, VI, VII, VIII, XI, and XV), alongside several metastable phases~\cite{petrenko1999physics,salzmann2019advances} and new polymorphs reported in the last few years, such as ice XIX, ice XXI, and plastic ice VII ~\cite{lee2025multiple,gasser2021structural,yamane2021experimental,rescigno2025observation}.
In many ice structures, the oxygen atoms arrange into a well-defined crystalline lattice, while the hydrogen atoms remain ordered or disordered, reflecting the near-degenerate energies of different proton configurations permitted by the ice rules~\cite{bernal1933theory,pauling1935structure}.
The corresponding residual entropy~\cite{giauque1936entropy,pauling1935structure} stabilizes disordered polymorphs at higher temperatures, 
whereas upon cooling, ordered phases, such as ice XI, become favored~\cite{salzmann2019advances}.
Accurately describing this complex interplay between bonding, entropy, and proton order–disorder transitions remains a non-trivial test for atomistic models.

Despite the conceptual simplicity of the H$_2$O molecule, modeling its condensed phases requires careful compromises between accuracy and computational feasibility~\cite{gillan2016perspective,cisneros2016modeling,brini2017water}.
Classical empirical models such as TIP4P/2005 and SPC/E reproduce selected thermodynamic properties but often lack predictive power beyond their fitted state points~\cite{jorgensen1983comparison,abascal2005general,vega2011simulating}.
Coarse-grained models like mW enable large-scale simulations but sacrifice molecular detail and cannot capture proton ordering effects~\cite{molinero2009water}.
First-principles molecular dynamics based on density functional theory (DFT) provide a more transferable and physically grounded description~\cite{gillan2016perspective,tuckerman1995ab}, yet their high computational cost restricts the accessible time and length scales required for determining the phase diagram.
Machine-learning interatomic potentials have emerged as a transformative approach, combining near \textit{ab initio} accuracy with orders-of-magnitude greater efficiency~\cite{bartok2017machine,deringer2021gaussian}. 
Frameworks such as the Deep Potential method~\cite{zhang2018deep} and Behler-Parrinello Neural Network Potentials~\cite{behler2007generalized} have demonstrated remarkable success in describing water’s complex behavior~\cite{cheng2019ab,piaggi2021phase,lan2021simulating, daru2022coupled, piaggi2022homogeneous, kapil2024first}.

\begin{figure*}[hbt!]
    \centering
    \includegraphics[width=\textwidth]{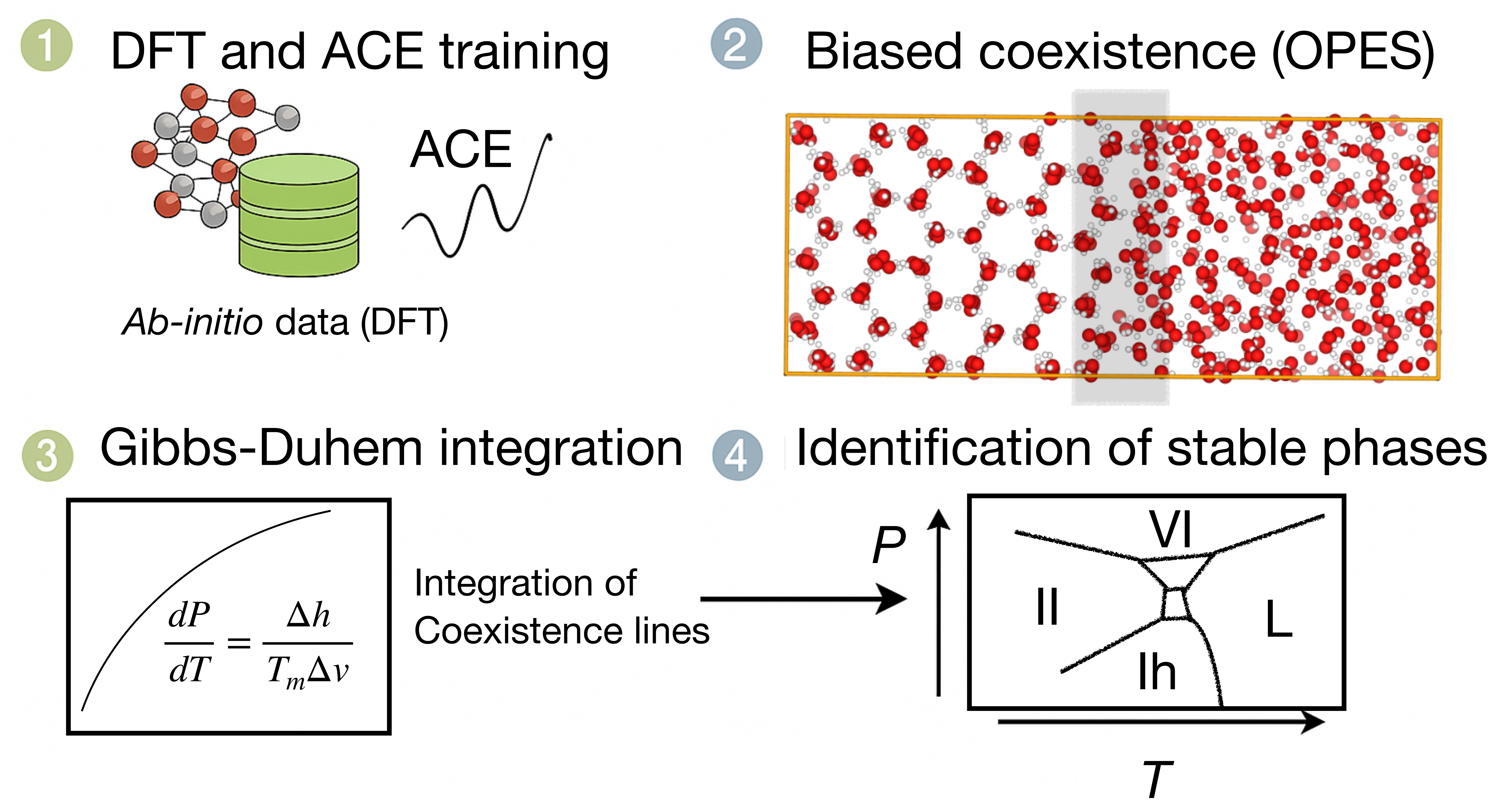}
    \caption{Schematic overview of the computational workflow employed in this study, combining \textit{ab initio} data, Atomic Cluster Expansion (ACE) training, biased coexistence simulations with OPES, and Gibbs–Duhem integration to construct the phase diagram of water.}
    \label{fig:workflow}
\end{figure*}

We recently introduced a transferable Atomic Cluster Expansion (ACE) potential for water trained on data of various ice phases computed using DFT with the revPBE-D3 exchange and correlation functional~\cite{ibrahim2024efficient}. 
Our model accurately reproduces liquid water properties including radial distribution functions, diffusion coefficient, and vibrational spectra, as well as energy-volume curves of various ice polymorphs, demonstrating excellent transferability and robustness of the ACE representation. 
Starting from this ACE potential, we supplement our previous training set with additional data in order to be able to cover a wide range of temperatures and pressures on the water phase diagram.
To determine the phase boundaries with high accuracy, we first use the biased coexistence method \cite{bore2022phase} based on the On-the-fly Probability Enhanced Sampling (OPES technique~\cite{invernizzi2020rethinking,invernizzi2020unified}, which enables the calculation of liquid-solid chemical potential differences and melting points.
Afterwards, we employ Gibbs–Duhem integration\cite{kofke1993direct} for tracing coexistence lines consistently across temperature and pressure.
In Figure~\ref{fig:workflow}, we illustrate the computational workflow that we employ in this study.
The resulting phase diagram exhibits good agreement with experimental boundaries~\cite{salzmann2011polymorphism}, as well as with previous models trained on data derived from DFT~\cite{zhang2021phase} or from the MB-pol model~\cite{bore2023realistic}, which is based on CCSD(T) coupled-cluster calculations.

\section{Methods}
\label{sec:methods}

\subsection{ACE Potential and Training Procedure}

In order to train a transferable model for water and ice, here we employ the ACE formalism.
This methodology provides a formally complete descriptor of the local atomic environments for a systematically convergent and physically interpretable representation of interatomic interactions~\cite{drautz2019atomic,dusson2022atomic}. 
In particular, the local atomic energy is expressed as a body-ordered expansion over invariant basis functions, which can be truncated at any desired body order. 
We employed a Finnis–Sinclair embedding with non-linear representation of the atomic energy that incorporates two atomic properties, each represented by ACE basis expansions~\cite{drautz2019atomic,lysogorskiy2021performant,bochkarev2022efficient}.
This non-linear form, originally motivated by the second-moment approximation, has proven to be efficient for both metallic~\cite{bochkarev2022efficient,lysogorskiy2021performant,e_ibrahim_magnesium} and covalently bonded materials~\cite{qamar2023atomic}. Non-linear representations were also beneficial for multi-layer ACE~\cite{bochkarev2022multilayer}, multi-component materials~\cite{ibrahim2024efficient,bienvenu2025development,starikov2025atomic} as well as for systems with non-collinear magnetic degrees of freedom~\cite{rinaldi2024non}, ACE with charge~\cite{rinaldi2025charge}, ACE-based graph representations~\cite{bochkarev2024graph}, and graph-based ACE for foundation models~\cite{lysogorskiy2025graph}.

We start from our previously developed dataset based on DFT with the revPBE-D3 exchange and correlation functional~\cite{ibrahim2024efficient}.
The original dataset contains diverse configurations of multiple ice structures, both proton-ordered and proton-disordered, covering densities between 0.9\, and 1.8\,g\,cm$^{-3}$.  
Our previous ACE potential showed an excellent description of the potential energy surface (PES) for diverse ice phases as well as for liquid water.
Only a few liquid water configurations were included in the dataset using an active learning strategy to improve the description of liquid water without the need to run computationally expensive \textit{ab initio} molecular dynamics (AIMD). 
The ACE potential achieved high transferability and accuracy across both solid and liquid phases, reproducing structural, dynamical, and thermodynamic properties in close agreement with DFT and experiment at ambient conditions.

To extend the applicability of the previously published ACE potential to high-pressure relevant for the present phase diagram calculations, we included a small number of configurations from the dataset of the SCAN-trained DeepMD potential water model~\cite{zhang2021phase}.
The dataset was processed and the extrapolation grade of our previous ACE model for each configuration was evaluated based on the D-optimality criterion~\cite{lysogorskiy2023active,podryabinkin2017active}.
Then, configurations with high extrapolation grades, primarily dense ice polymorphs and compressed liquid-like environments, were extracted and their energies and forces were recalculated with DFT using the revPBE-D3 functional, consistent with the setup of our previously published ACE potential~\cite{ibrahim2024efficient}.
The resulting extension, involving fewer than one hundred additional configurations, effectively removed extrapolation across the entire pressure–temperature range, enhancing numerical stability and accuracy while preserving the model’s transferability.
The resulting dataset contains only 2,949\, configurations in total, while the previously published DeepMD-SCAN training set comprises 31,058\, configurations.

All ACE parameterizations were performed using the PACEmaker package~\cite{bochkarev2022efficient}, employing a hierarchical basis extension scheme with power-order ranking of basis functions. 
The expansion was truncated at fifth body order with a cutoff radius of 6.0\,\AA.
Energies and forces were weighted in the loss function using an energy-based weighting scheme that assigns higher weights to low-energy configurations near equilibrium.

\subsection{OPES Enhanced Sampling Coexistence Simulations}
Solid-liquid coexistence points were computed using biased coexistence simulations based on the OPES method, employing the expanded-ensemble variant that targets a flat distribution of the chosen collective variables (CVs)~\cite{invernizzi2020unified}.
This approach enables the reversible growth and melt of a layer of solid within a single trajectory, allowing the direct estimation of the solid-liquid chemical potential difference~\cite{bore2022phase} and is similar to the interface pinning method~\cite{pedersen2013computing}.

Initial solid-liquid coexistence configurations were constructed following the protocol of Bore \textit{et al.}~\cite{bore2022phase}. 
First, each ice polymorph was equilibrated for 1\,ns in the NPT ensemble using the ACE potential. 
The equilibrium box dimensions were averaged over this trajectory and subsequently used to define a fixed cross-sectional area for the coexistence setup.
A slab of equilibrated ice was then duplicated along its longest dimension to create a configuration containing a frozen half and a mobile half. 
The mobile region was subjected to heating until fully melted, after which the full system was re-equilibrated at the target thermodynamic conditions.
All enhanced-sampling simulations were performed in the NPT ensemble with flexible box lengths along the solid-liquid interface normal and fixed lateral dimensions.

The CVs were constructed using the environment similarity kernel that quantifies the degree to which the oxygen coordination environment matches that of a reference ice structure~\cite{piaggi2019calculation}.
We consider the atomic environments within a prescribed cutoff $r_c$ of each water molecule in a given ice polymorph.
Each polymorph can be described by $X=\chi_1,....,\chi_m$ distinct local environments.
The environment similarity kernel that compares the atomic environments $\chi_l \in X$ with a generic environment $\chi$ is obtained as,
\begin{equation}
    k_{\chi_l}(\chi) = \frac{1}{n} \sum_{i \in \chi_l} \sum_{j \in \chi} \exp\left(-\frac{|\mathbf{r}_i^l-\mathbf{r}_j|^2}{4 \sigma^2}\right) \,,
    \label{eq:kernel2}
\end{equation}
where $n$ is the number of neighbors in the environment $\chi_l$, $\sigma$ is a broadening parameter, and $\mathbf{r}_i^l$ and $\mathbf{r}_j$ are the positions of the neighbors in environments $\chi_l$ and $\chi$, respectively.
We then define a best-match kernel $k_X(\chi)$, which is a single similarity measure between a given environment and any of the $m$ reference environments of a polymorph,
\begin{equation}
    k_X(\chi) = \frac{1}{\lambda} \log 
    \left (
    \sum\limits_{l=1}^{m} \exp (\lambda k_{\chi_l} ) \,.
    \right)
    \label{eq:kernel_multi2}
\end{equation}
The parameter $\lambda$ controls the smoothness of the function and was set to $100$.
Finally, using this similarity metric we define the total number of ice-like environments in the simulation box as,
\begin{equation}
    N_{\mathrm{ice}} = \sum\limits_{i=1}^N f(k_X(\chi^i))
    \label{eq:op_num_2}
\end{equation}
where $N$ is the total number of oxygen atoms, $\chi^i$ is the atomic environment around the $i$-th oxygen atom, and $f$ is a switching function that is $\sim0$ and $\sim1$ for $k_X(\chi^i)$ values consistent with the liquid and the solid, respectively. We take
\begin{equation}
f(y)=
\begin{cases}
    0 \quad \mathrm{if} \quad y<0 \\
    y^2(3-2y) \quad \mathrm{if} \quad 0<y<1 \\
    1 \quad \mathrm{if} \quad y>1\\
\end{cases} \,,
\end{equation}
where $y=(k_X(\chi^i)-k_1)/(k_2-k_1)$, and $k_1$ and $k_2$ are chosen as the peaks of the liquid and ice distributions of $k_X(\chi^i)$, respectively.
We further can define a strict CV, instead of a smooth one, using a threshold $\kappa=(k_1+k_2)/2$ above which an environment is considered to be ice-like.

For each ice phase, the environments were identified and a corresponding CV $N_{\mathrm{ice}}$ was defined.
We employed the same parameters used in Ref.~\citenum{bore2022phase} for ice Ih, II, III, V, and VI.
In addition, we also constructed a suitable CV $N_{\mathrm{ice}}$ for ice VII by choosing an appropriate Gaussian width $\sigma$, cutoff radius $r_c$, and solid-liquid thresholds $k_1$ and $k_2$ to ensure minimal overlap between the liquid and ice-VII distributions.
The chosen values are summarized in Table.~\ref{table:cvparams}.

\begin{table}[hbt!]
\begin{center}
  \begin{tabular}{l|c}
  property & value \\
  \hline
  $m$ & 1 \\
  $r_c$ [nm] &  0.34 \\
  $\sigma$ [nm] & 0.0700 \\
  $k_1$ & 0.69 \\
  $k_2$ & 1.05625 \\
  $\kappa$ & 0.8731 \\
  $N_{\mathrm{H_2O}}$ & 720 \\
  Box dimensions [nm$^{3}$] & 20.21 $\times$ 20.26 $\times$ 36.18 \\
  $N_{\mathrm{ice}}$ range & 360--432 \\
  \end{tabular}
\end{center}
\caption{Parameters used in defining the environment-similarity collective variable for the ice~VII polymorph and simulation setup for the OPES coexistence simulations.}
\label{table:cvparams}
\end{table}

\begin{figure*}[hbt!]
    \centering
    \includegraphics[width=\textwidth]{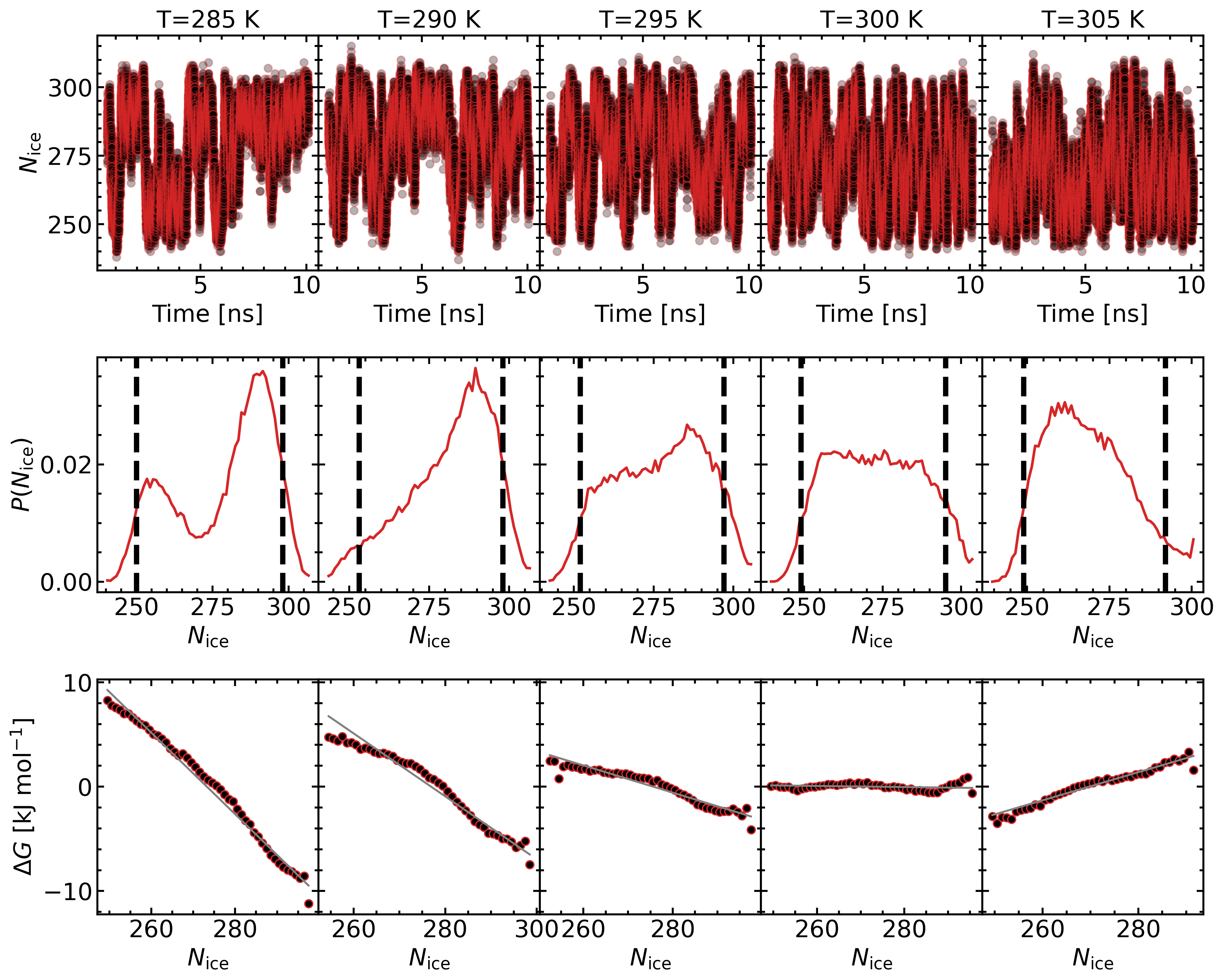}
    \caption{Representative OPES biased-coexistence simulations used to determine solid-liquid free-energy differences for ice Ih polymorph at 0.1\,GPa. See text for details.
    (Top row) Time series of the collective variable $N_{\mathrm{ice}}$ at selected temperatures, showing repeated and reversible growth and melting events of an ice layer. (Middle row) Biased probability distributions of $N_{\mathrm{ice}}$. (Bottom row) Corresponding reweighted free-energy profiles $\Delta G(N_{\mathrm{ice}})$ together with linear fits used to extract the chemical-potential difference $\Delta\mu = \mu_{\mathrm{ice}} - \mu_{\mathrm{liq}}$.
    }
    \label{fig:opes}
\end{figure*}

In order to compute ice-liquid chemical potential differences and melting points, we constructed a bias potential based on OPES expanded \cite{invernizzi2020unified,bore2022phase} using $N_{\mathrm{ice}}$ as collective variable.
Once the bias potential produced by OPES had converged, the unbiased free-energy profile $ \Delta G(N_{\mathrm{ice}})$ as a function of the number of ice-like molecules, $N_{\mathrm{ice}}$, was reconstructed by means of reweighting \cite{bore2022phase}.

It can be shown that the chemical potential difference $\Delta\mu = \mu_{\mathrm{ice}} - \mu_{\mathrm{liq}}$ is connected to $ \Delta G(N_{\mathrm{ice}})$ through,
\begin{equation}
    \frac{
    \Delta G(N_{\mathrm{ice}} + \Delta n_{\mathrm{layer}}) -
    \Delta G(N_{\mathrm{ice}})
    }{
    \Delta n_{\mathrm{layer}}
    }
    =
    \Delta\mu,
    \label{eq:chemical_potential_difference}
\end{equation}
where $\Delta n_{\mathrm{layer}}$ is the number of molecules in the layer of ice that reversibly grows and melts during the coexistence simulation.  

In practice, $\Delta G(N_{\mathrm{ice}})$ was fit with an error-weighted linear regression over the interval corresponding to the reversible formation or melting of a single crystalline layer, and the slope is $\Delta\mu$. Repeating this procedure at several temperatures along the same isobar allowed us to extract $\Delta\mu(T)$, from which the melting temperature $T_{\mathrm{eq}}$ was identified by the condition 
    $\Delta\mu(T_{\mathrm{eq}}) = 0$.

In Figure~\ref{fig:opes}, we show representative OPES-based coexistence simulations that we used in this work to compute $\Delta\mu(T)$ for the ice polymorph Ih at 0.1\,GPa.
In the top row, we report the time evolution of the collective variable $N_{\mathrm{ice}}$, which shows the number of ice-like molecules. 
We observe that during the simulation a layer of ice reversibly grows and melts.
In the middle row, we compute and display the biased probability distributions $P(N_{\mathrm{ice}})$ computed from the trajectories in the top row.
As expected, the distributions are approximately flat consistent with the target distribution in OPES expanded.
In the bottom row, we reconstruct the free-energy profiles $\Delta G(N_{\mathrm{ice}})$ using reweighting and extract the solid-liquid chemical-potential differences $\Delta\mu$ from the slope of linear fits to the data.
This protocol provides statistically robust estimates of melting points without relying on assumptions about proton disorder or separate free-energy references for liquid and solid.

\subsection{Gibbs–Duhem Integration}
Coexistence lines were computed across pressure and temperature using Gibbs-Duhem integration starting from the OPES-determined coexistence points. 
The Clapeyron relation,
\begin{equation}
\frac{{\mathrm d}P}{{\mathrm d}T} = \frac{\Delta h}{T\,\Delta v},
\end{equation}
was integrated numerically, where $\Delta h$ and $\Delta v$ denote per-molecule enthalpy and volume differences between the two phases. These quantities were obtained from equilibrium NPT simulations performed at each integration step.
Depending on the slope of the coexistence curve, either the pressure or temperature was used as the integration variable.
Pressure-based integrations employed increments of 0.0125\,GPa (125\,bar), while temperature-based integrations used steps of 3\,K. For each state point, ice and liquid phases were equilibrated for 10\,ps followed by 100\,ps of production, whereas liquid water was simulated for 1\,ns.
Integration was done using the fourth-order Runge-Kutta method~\cite{dormand1980family}.

This procedure enables tracing of coexistence lines without additional free-energy calculations and ensures thermodynamic consistency across the full range of thermodynamic conditions examined in this study.

\subsection{Reference calculations}

The DFT reference calculations for the configurations identified through active learning were performed using the Vienna Ab initio Simulation Package (VASP)\cite{kresse1996efficiency,kresse1996efficient}. 
The electronic structure was described within the framework of the Perdew–Burke–Ernzerhof (revPBE) generalized gradient approximation (GGA)\cite{perdew1996generalized,perdew1998perdew}. 
To account for long-range dispersion interactions, the Grimme D3 correction with Becke-Johnson damping was employed~\cite{grimme2010consistent,grimme2011effect}. 
This approach has been shown to partially offset intrinsic functional errors in PBE~\cite{pestana2017ab}, yielding a balanced and experimentally consistent description of condensed-phase water in classical nuclear dynamics.
A plane-wave energy cutoff of 450\,eV and k-mesh density of 0.125\,\AA$^{-1}$ were used.

\subsection{Simulation Details}
All molecular dynamics simulations were performed using the LAMMPS package~\cite{LAMMPS} with the PACE pair-style implementation of ACE.
Periodic boundary conditions were applied in all directions.
Equilibration of the ice boxes are done for 1\,ns for the determination of two-phase interface area.
All coexistence simulations were carried out using LAMMPS~\cite{LAMMPS} interfaced with PLUMED~2.9.2~\cite{tribello2014plumed,plumed2019promoting}.
The total simulation length for each simulation was 10\,ns, with an initial equilibration time of 100\,ps.

\section{Results}
\label{sec:results}

\subsection{Accuracy and Transferability of the ACE Potential}

\begin{figure*}[hbt!]
    \centering
    \includegraphics[width=\textwidth]{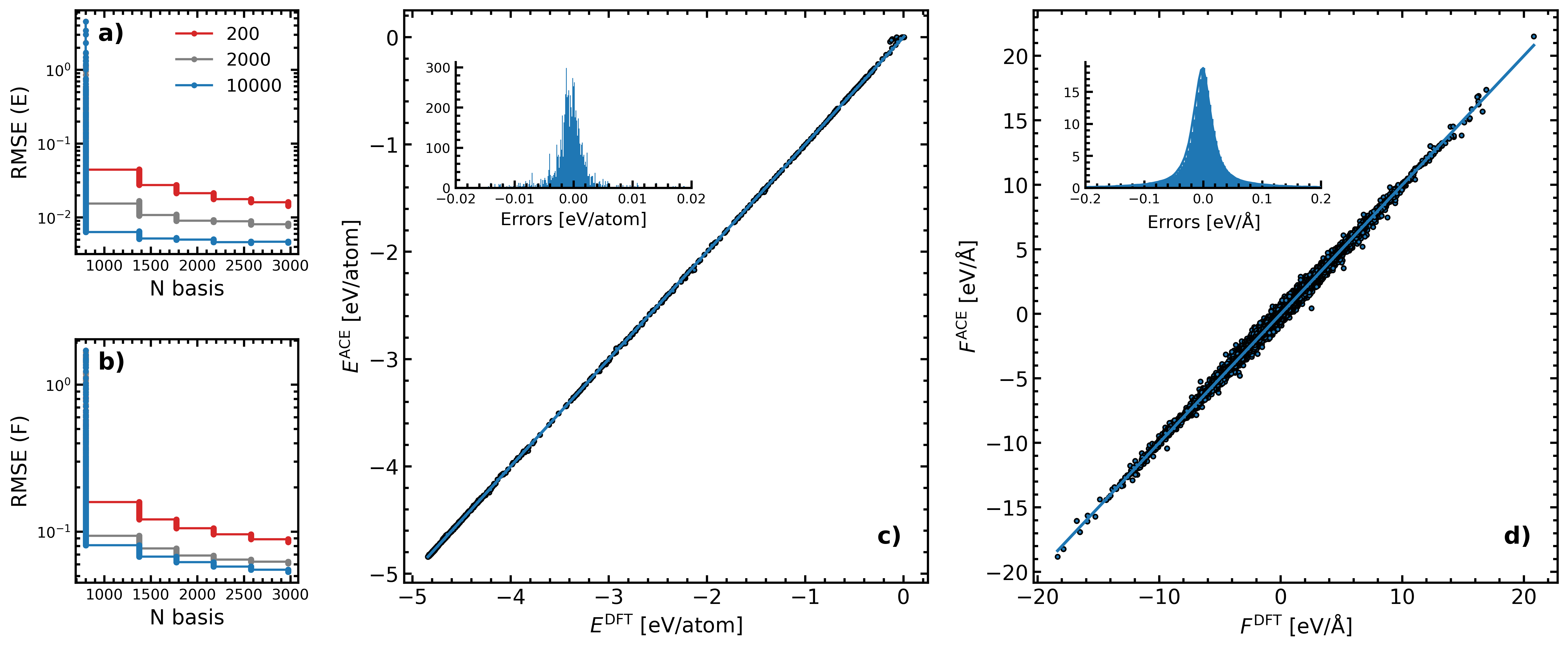}
    \caption{ACE potential training. Panels a and b show the convergence of the root-mean-square error (RMSE) of energies and forces, respectively, as a function of the number of basis functions for different number of optimization steps. Panels c and d show parity plots between the DFT reference data and the ACE predictions for energies and  forces, respectively. Insets show the distribution of errors. 
    }
    \label{fig:train_data}
\end{figure*}

Figures~\ref{fig:train_data}a and \ref{fig:train_data}b illustrate the convergence of the RMSE of the energies and forces, respectively, with respect to the number of ACE basis functions and the number of optimization steps.
We see that for a fixed basis size, increasing the number of optimization steps systematically reduces the RMSE, as shown by the red (200 steps), gray (2,000 steps), and blue (10,000 steps) curves.
This behavior demonstrates that additional optimization improves accuracy without increasing the complexity of the basis.
Based on this analysis, we selected a compact ACE representation consisting of 1,374 basis functions (3,942 parameters) optimized using 10,000 steps.
This final model retains the accuracy of our previously published ACE potential~\cite{ibrahim2024efficient}, while using a smaller basis and an extended dataset that covers a broader pressure–temperature range.

In Figures~\ref{fig:train_data}c and \ref{fig:train_data}d we show the accuracy of the final ACE potential by comparing predicted energies and forces with the DFT reference data.
The parity plots exhibit a narrow scatter around the diagonal, demonstrating that the ACE model faithfully reproduces the underlying first-principles potential energy surface.
The insets provide a quantitative assessment of the error distribution.
Both the energy and force error distributions are sharply peaked near zero and decay smoothly, indicating that the errors are well controlled and that no significant outliers are present.

To assess transferability beyond the training domain, we evaluated the full SCAN dataset comprising 31,058 configurations~\cite{zhang2021phase} using the D-optimality criterion~\cite{lysogorskiy2023active,podryabinkin2017active}.
Although only about 100 configurations from this dataset were incorporated into the training set, none of the remaining configurations were identified as extrapolative with respect to the final ACE model.
We applied the same criterion to all simulations used for the phase diagram calculations and likewise detected no extrapolation.

These results demonstrate that the final ACE potential provides an accurate and transferable description of water across the full range of thermodynamic conditions explored in this work.

\subsection{Thermodynamic Stability and Chemical Potentials}
\begin{figure*}[hbt!]
    \centering
    \includegraphics[width=18cm]{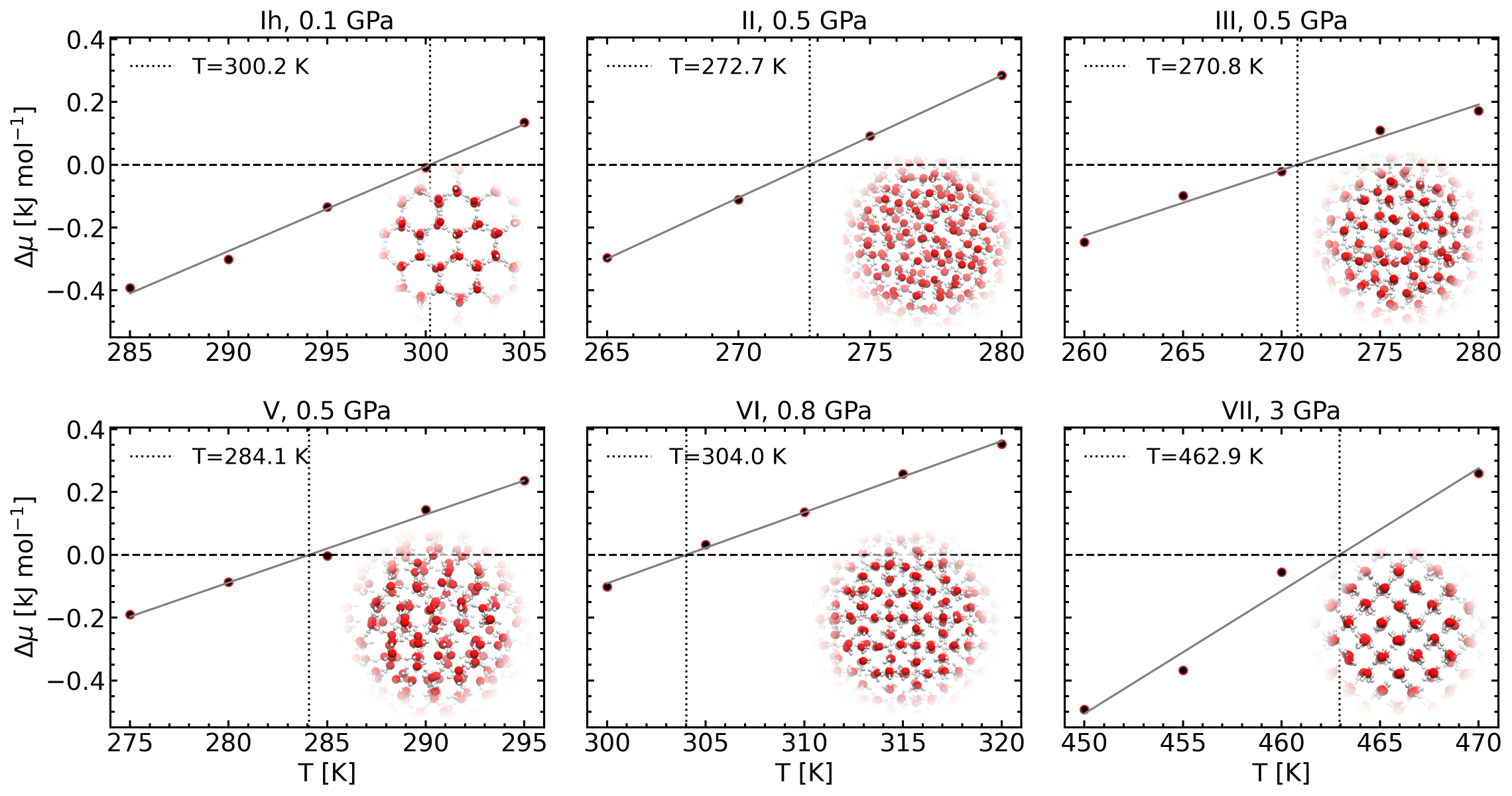}
    \caption{Chemical-potential differences for ice polymorphs relative to the liquid as a function of temperature at fixed pressure. The horizontal dashed line marks $\Delta\mu = 0$; vertical dotted lines indicate the melting point.
}
    \label{fig:chem_pots}
\end{figure*}

The thermodynamic stability of the various ice polymorphs relative to the liquid phase was assessed by computing the chemical-potential differences $\Delta\mu = \mu_{\mathrm{ice}} - \mu_{\mathrm{liq}}$ using OPES-based biased coexistence simulations, as outlined in the methods section.
In Figure~\ref{fig:chem_pots}, we show $\Delta\mu$ for multiple temperatures around the melting point and at a fixed pressure for each ice polymorph.
The resulting $\Delta\mu(T)$ curves exhibit a linear trend in the studied temperature range for all ice polymorphs, as expected from the weak dependence of entropy with temperature.
We computed the coexistence temperature $T_{\mathrm{eq}}$ by performing a linear fit to $\Delta\mu(T)$ and finding the temperature for which $\Delta\mu = 0$.
We show the coexistence temperature in Figure~\ref{fig:chem_pots} using vertical dotted lines and we estimate the error in the determination of coexistence temperatures to be below 1\,K.
The small scatter of the data points around the fitted lines illustrates the smooth free-energy landscape produced by the ACE potential and the  statistical reliability of the OPES sampling and reweighting procedure.

\subsection{Gibbs–Duhem Integration and Coexistence Lines}

\begin{figure}[hbt!]
    \centering
    \includegraphics[width=\columnwidth]{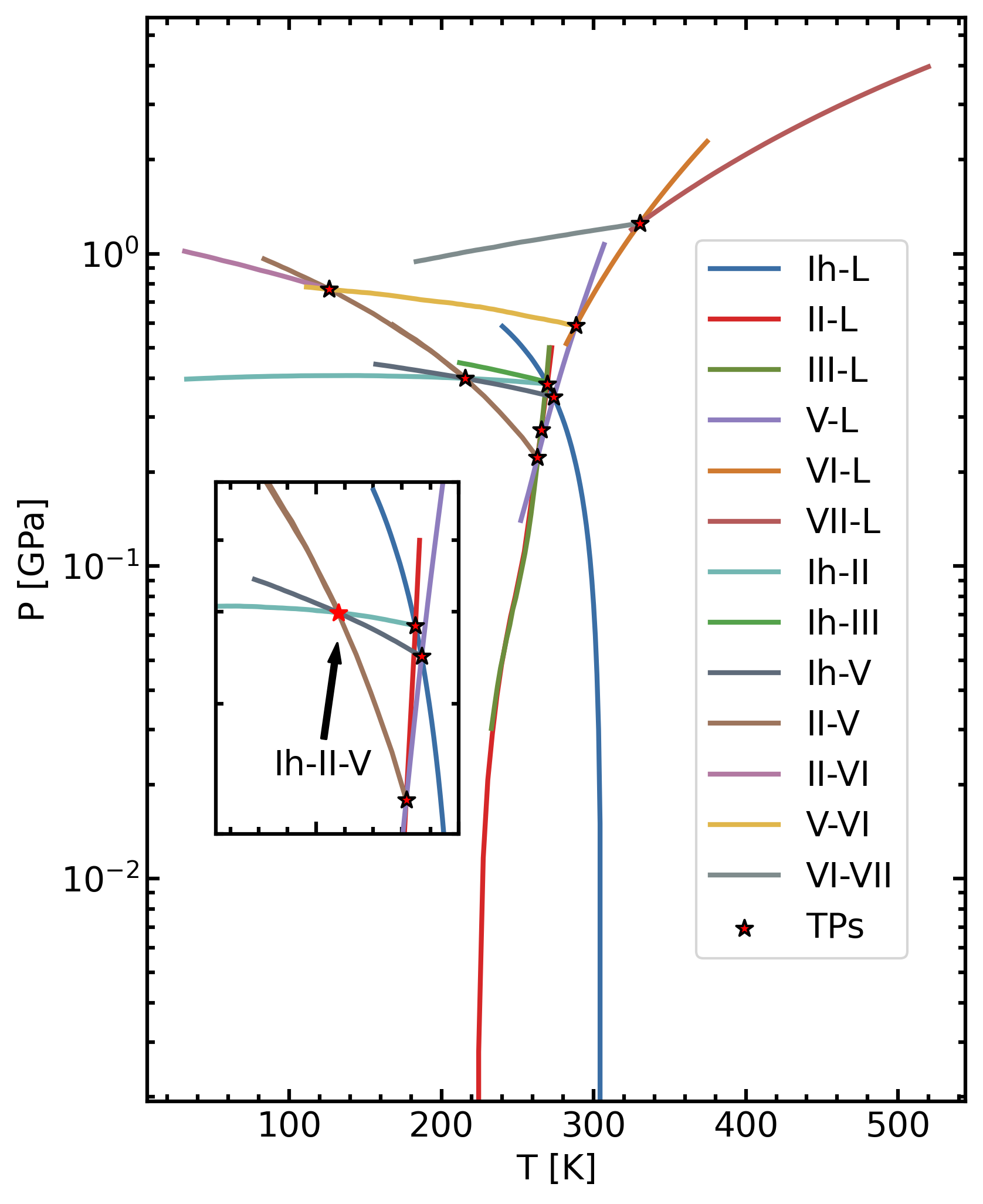}
    \caption{Gibbs--Duhem integration of phase coexistence lines.
Each curve represents a solid--liquid or solid--solid coexistence boundary obtained by integrating from coexistence points obtained using OPES simulations.
Colored lines correspond to coexistence lines between the  phases indicated in the legend (L for liquid, and ice polymorphs Ih, II, III, V, VI, and VII), while stars denote triple points (TPs) identified from the intersections of coexistence lines.
The inset highlights the region where we tested the robustness of our calculation by computing the Ih-II-V triple points using different routes.}
    \label{fig:gibbs_duhem}
\end{figure}

Figure~\ref{fig:gibbs_duhem} shows the coexistence lines that we obtained from Gibbs–Duhem integration.
We initiated each numerical integration from the melting points computed in the previous section with OPES-based biased coexistence simulations.
From these points we propagated the solid–liquid phase boundaries to cover the desired temperature-pressure range.
Once the liquid-solid phase boundaries were computed, we determined the liquid-solid-solid triple points as the points where solid-liquid phase boundaries of two ice polymorphs meet.
Starting from these triple points, we performed new Gibbs-Duhem integrations to determine solid-solid phase boundaries.
As we show in Figure~\ref{fig:gibbs_duhem}, the integrated coexistence curves remain smooth and continuous, and the slopes match the trends reported in experiments and previous theoretical work~\cite{zhang2021phase,bore2023realistic}.
We did not observe discontinuities or problems during the numerical integration procedure, which indicates that the ACE potential provides stable and accurate dynamics across the entire pressure–temperature range considered.

We performed a consistency test for the triple point involving ice Ih, II, and V, based on determining it  using different routes.
For this purpose, we integrated the Ih–V, Ih-II, and II-V coexistence lines starting from their respective liquid-solid-solid triple points.
The intersection of any two of these three lines determines the ice Ih-II-V triple point.
We found that the three coexistence lines converge to the same state point within numerical uncertainty (0.1\,K and $1.4\times10^{-5}$\,GPa).
The inset of Figure~\ref{fig:gibbs_duhem} highlights the relevant region, where the Ih–V, II–V, and II–Ih lines meet at a common point.
This test demonstrates the reliability of the OPES coexistence data and the accuracy of the Gibbs–Duhem integration, and supports the overall consistency of the reconstructed phase diagram.

\subsection{Water Phase Diagram}
\begin{figure*}[hbt!]
    \centering
    \includegraphics[width=\textwidth]{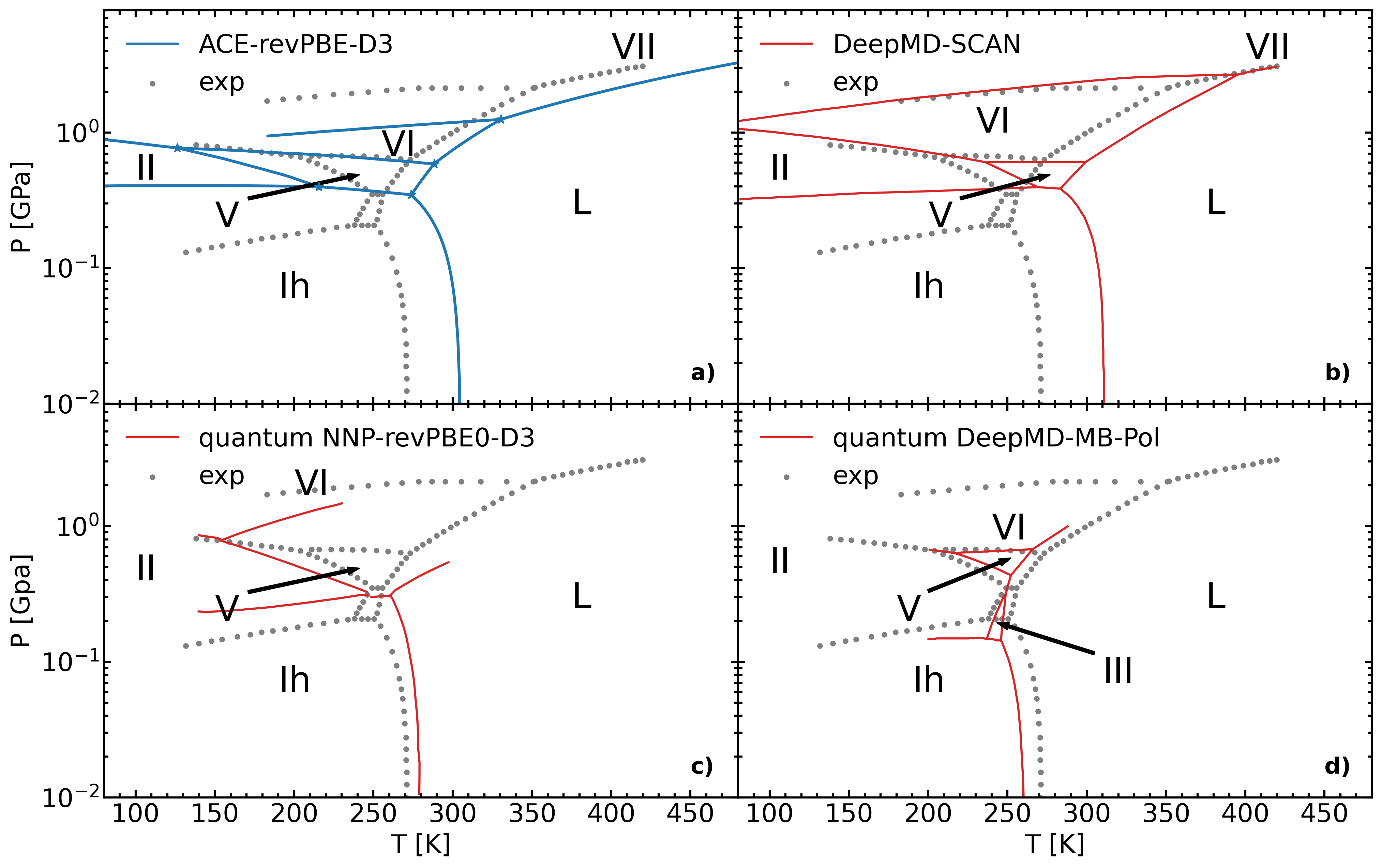}
    \caption{
Computed phase diagram of water compared with experimental data~\cite{salzmann2019advances} and several state-of-the-art machine learning models.
Panel~(a) shows results obtained with the ACE-revPBE-D3 potential, while panels~(b)-(d) correspond to DeepMD-SCAN~\cite{zhang2021phase}, NNP-revPBE0-D3~\cite{reinhardt2021quantum}, and MB-pol~\cite{bore2023realistic}, respectively.
The latter two models include nuclear quantum effects as well as corrections through free-energy perturbation.
Experimental phase boundaries are shown in gray\citenum{salzmann2019advances}.
Labels indicate the stable solid polymorphs and the liquid (L) region.
}
    \label{fig:pd}
\end{figure*}

In Figure~\ref{fig:pd}a, we show the complete pressure–temperature phase diagram for our ACE potential.
In order to construct this phase diagram, we first identified the stable triple points.
These points were computed from the intersection of stable phase boundary lines and they are summarized in Table~\ref{table:triple_points}.
Afterwards, we joined these triple points with the corresponding phase coexistence lines and we removed the part of the line beyond the stable triple points.
For example, the ice VI-liquid phase boundary is obtained by joining the V-VI-liquid and VI-VII-liquid triple points.
The predicted phase boundaries for the main crystalline polymorphs (Ih, II, V, VI, and VII) agree qualitatively with experimental results~\cite{salzmann2019advances}, also shown in Figure~\ref{fig:pd}a.
Most ice polymorphs are positioned in the correct region of the phase diagram, except ice III which should be stable, and is only metastable.
The slopes of the phase boundaries agree very well with experiment, reflecting that our potential correctly captures the changes in density and entropy with temperature.
However, there are systematic shifts, such as for the Ih-liquid phase boundary, which is displaced to higher temperatures by around 30\,K.
As we shall discuss later, these limitations are common in models based on semi-local DFT. 
Thus, our ACE potential based on revPBE-D3 provides a reliable description of water thermodynamics across a wide range of pressures and temperatures.

\begin{table}[hbt!]
\begin{center}
  \begin{tabular}{l|cc}
  Triple point & $P$ [GPa] & $T$ [K]\\
  \hline
  Liquid--Ih--V   & 0.348 & 274 \\
  Liquid--V--VI   & 0.589 & 288 \\
  Liquid--VI--VII & 1.251 & 330 \\
  Ih--II--V       & 0.399 & 216 \\
  II--V--VI       & 0.769 & 126 \\
  \end{tabular}
\end{center}
\caption{Triple points of the ACE water model obtained from the intersections of coexistence lines, which is reconstructed with Gibbs-Duhem integration.}
\label{table:triple_points}
\end{table}

We next compare the ACE phase diagram with results obtained from other state-of-the-art machine-learning water models.
Figure~\ref{fig:pd}b shows the phase diagram predicted by the DeepMD-SCAN model~\cite{zhang2021phase}, which was trained on energies and forces computed using the SCAN meta-GGA functional~\cite{sun2015strongly}.
The overall topology of the DeepMD-SCAN phase diagram is similar to that obtained with the ACE-revPBE-D3 potential.
In particular, both models predict ice~III to be metastable and exhibit systematic shifts of the solid--liquid coexistence lines toward higher temperatures relative to experiment.
For ice~Ih, the ACE-revPBE-D3 potential predicts a melting temperature of 305\,K at 1 bar, in somewhat better agreement with experiment (273 K) than DeepMD-SCAN (310 K).
Similar trends are observed for the VI-liquid and VII-liquid coexistence lines, where the ACE-revPBE-D3 predictions lie closer to the experimental melting boundaries than those obtained with DeepMD-SCAN.
Differences between the two models are also evident in the stability region of ice~V.
While the extent of the ice~V stability region predicted by DeepMD-SCAN is closer to experiment, the slope and location of the ice~V melting line are reproduced more accurately by the ACE-revPBE-D3 potential.

In Figure~\ref{fig:pd}c, we compare our results with the water phase diagram obtained using the NNP-revPBE0-D3 model~\cite{reinhardt2021quantum}, which is a Behler-Parrinello neural network potential trained on revPBE0-D3 reference data~\cite{cheng2019ab}.
The revPBE0-D3 functional is a hybrid exchange-correlation functional that includes a fraction of exact exchange and therefore represents a higher level of electronic-structure theory than the semi-local functionals discussed above.
In addition, the NNP-revPBE0-D3 phase diagram incorporates both free-energy perturbation corrections to the underlying DFT reference and explicit nuclear quantum effects through path-integral simulations~\cite{reinhardt2021quantum}.
This NNP-revPBE0-D3 model provides a substantially improved prediction of the ice~Ih melting line compared to both the ACE-revPBE-D3 and DeepMD-SCAN models.
At the same time, notable discrepancies exist for other phase boundaries.
In particular, the slopes of the V-liquid and V-VI coexistence lines differ significantly from experimental observations and from the trends predicted by the other models discussed above.
Moreover, the stability region of ice~V is considerably larger than its experimental counterpart.
Further work is needed to elucidate the reasons why the inclusion of exact exchange, i.e., going from revPBE-D3 to revPBE0-D3, does not seem to lead to a systematic improvement of the phase diagram.

Finally, we compare our results with the phase diagram obtained using the MB-pol model~\cite{babin2014development,bore2023realistic}, shown in Figure~\ref{fig:pd}d.
MB-pol is a many-body potential explicitly constructed from coupled-cluster reference data, combining accurate two-body and three-body interaction terms with a classical polarization model for higher-order contributions.
The phase diagram shown in Figure~\ref{fig:pd}d was obtained using a DeepMD potential trained to reproduce MB-pol energies and forces~\cite{bore2023realistic} and for this reason we refer to this model as DeePMD-MB-Pol.
The calculations also incorporate free-energy perturbation corrections and explicit nuclear quantum effects through path-integral simulations.
As shown in Figure~\ref{fig:pd}d, the phase diagram based on DeePMD-MB-Pol provides a better overall description of water’s phase behavior and is the only model discussed here that predicts ice~III to be thermodynamically stable.
Moreover, the locations of most phase boundaries are in closer agreement with experiment than those obtained from the other models considered above.

We suggest that the deviations observed in our revPBE-D3-based ACE phase diagram should be attributed mainly to the limitations of the underlying first-principles reference and, to a lesser extent, to the neglect of nuclear quantum effects.
In contrast, we do not expect that errors arising the ACE formalism play a significant role in the deviations observed in the phase diagram, considering the high accuracy of the model and the absence of extrapolative behavior across all simulations reported here.
Owing to its formally complete and systematically improvable representation, the ACE framework is expected to achieve accuracy comparable to that of MB-pol when trained on higher-level electronic-structure data.
At the same time, unlike MB-pol, ACE potentials are fully reactive and can describe bond breaking and proton transfer, enabling access to simulations beyond molecular water.

In summary, our general-purpose revPBE--D3-based ACE potential reproduces the experimental water phase diagram with good accuracy in comparison to the experiment and stat-of-the-art theoretical results.

\section{Conclusion}

In this work, we extended our recently published ACE potential for water based on the revPBE--D3 functional in order to create a general-purpose potential for all known ice polymorphs and liquid water across a broad range of pressures and temperatures.
We then computed the phase diagram of water by combining OPES-based coexistence simulations for calculating melting points with Gibbs–Duhem integration for propagating coexistence lines.
The resulting topology and coexistence boundaries agree well with experiment.
However, our potential fails to capture ice III to be stable, and the melting lines of multiple ice polymorphs are shifted to higher temperatures by around 20-30\,K.
These limitations are similar to those of other models based on DFT, such as that based on the SCAN functional.
Despite these limitations, our analysis shows that the general-purpose ACE potential captures with high accuracy the complex phase behavior of water in a very large region of the phase diagram and thus represents a valuable tool to explore a broad range of phenomena in water science. Moreover, our results demonstrate that the accuracy and efficiency of the ACE formalism in combination with state-of-the-art sampling methods allows for the calculation of complex phase diagrams from first principles.  
Future work will extend this potential to aqueous electrolytes and mineral–water interfaces, where developing accurate and efficient potentials remains a frontier of atomistic machine learning.

\section*{Data availability}
The final DFT reference dataset, ACE potential files, and associated data are available on GitHub and Zenodo~\cite{ibrahim_2026_18291736, eslam_ibrahim_2026_18291473}.

\section*{Acknowledgements}
E.I acknowledges funding through the International Max Planck Research School for Sustainable Metallurgy (IMPRS SusMet).
P.M.P. acknowledges funding from the Marie Skłodowska-Curie Cofund Programme of European Commission project H2020-MSCA-COFUND-2020-101034228-WOLFRAM2.
We acknowledge computational resources from the research center ZGH at Ruhr-University Bochum,  and Red Española de Supercomputación resources provided by Barcelona Supercomputing Center in MareNostrum 5 ACC to RES-FI-2024-2-0026 and FI-2024-3-0028.
We gratefully acknowledge the computing time provided on the high-performance computing system Noctua~2~\cite{noctua2} at the NHR Center Paderborn for Parallel Computing (PC$^2$), under project ID~4806.
The NHR Center PC$^2$ is jointly supported by the Federal Ministry of Education and Research and the state governments participating in the National High-Performance Computing (NHR) joint funding program.

\bibliography{refs.bib}
\end{document}